\documentclass[aps,prb,twocolumn]{revtex4-1} 

\usepackage{color}
\usepackage{amsmath}  
\usepackage{amsfonts} 
\usepackage{graphicx} 
\usepackage{hyperref}

\usepackage{times} 
\usepackage{bm}
\usepackage{slashed}

\newcommand{\pv}{\mathbf{p}}
\newcommand{\xv}{\mathbf{x}}

\newcommand{\red}[1]{{\color{red} #1}}
\renewcommand{\red}[1]{{ #1}}     % uncomment to get rid of red

\newcommand{\removefig}[1]{#1}

\begin{document}

\title{Active Learning in a Graduate Quantum Field Theory Course}

\author{G. Peter Lepage}
\email{g.p.lepage@cornell.edu} 
\affiliation{Department of Physics, Cornell University, Ithaca, NY, 14853}

\date{\today}

\begin{abstract}
This article describes how the author successfully adapted 
techniques drawn from the literature on active learning for 
use in a graduate-level course on quantum field theory. Students completed readings 
and online questions ahead of each class and spent class time working through problems that required them to practice the decisions and skills typical of a theoretical physicist. The instructor monitored these activities and regularly provided timely feedback to guide their thinking. Instructor-student interactions and student enthusiasm were similar to that encountered in one-on-one discussions with advanced graduate students.
Course coverage was not compromised. {The teaching techniques described here are well suited to other advanced courses.}
\end{abstract}

\maketitle

\section{Introduction}
Techniques drawn from the extensive research literature on active learning\cite{schwartz-book, nas, lang,cwsei}  are revolutionizing introductory courses in physics\cite{deslaurier} and other subjects.\cite{freeman} These techniques emphasize far more student-student and student-instructor interaction in the classroom than in more traditional formats, with much more fine-grained, real-time assessment (multiple times in each class) to help students evaluate their own understanding of the material, and to help instructors evaluate student understanding and provide targeted feedback. The goals of such instruction are less about the acquisition of facts, and more about ``deliberate practice'' of expert thinking and performance.\cite{wieman-daed, ericsson1, ericsson2}

Most physics education research has focused on lower-level courses, but research shows  that active-learning designs are also effective for upper-level courses, provided those designs are modified to reinforce the specific expert practices that underly the course material.\cite{optics} In this paper we describe how this approach was adapted by the author for a second-year graduate course on quantum field theory for particle physics. 

Active learning is well suited to graduate courses, where students must absorb a lot of technical material and learn how to manipulate it like an expert (since they aspire to become one). Much of what they must learn is about decisions: what to pay attention to, which tool to use, what strategy to adopt, how and when to test a potential solution, and so on. Active-learning designs give students far more practice making such decisions \red{than traditional lecture formats.} They also give instructors vastly more information about what their students are thinking, because the students are working carefully crafted problems in class, with the instructor wandering from student to student throughout. The instructor is likely to have direct interactions with every student in the class during each meeting time. The course becomes a long discussion between instructor and students\,---\,very similar in character to the one-on-one meetings Ph.D.\ advisors have with their advisees, at their office blackboards. This kind of coaching in expert thinking makes \red{the learning process} much more efficient and effective for students.

Active-learning designs also provide invaluable feedback to instructors about the actual obstacles faced by their students, as opposed to what the instructor imagines are \red{obstacles\,---\,there} is much more information than can be gleaned just from homework assignments and exams. This feedback allows the instructor to make real-time adjustments to the materials used inside and outside of class, based upon substantial amounts of data. It also means that active-learning courses can improve significantly after being taught once or twice.

Here we describe an implementation for active learning in a particular graduate-level course. {Our discussion is not intended as physics education research; rather it is designed as an example for colleagues who might wish to experiment with active learning in their advanced courses.} In Section~\ref{sec:course-structure}, we describe the goals and structure of our course. In Section~\ref{sec:activities}, we address the most challenging aspect for the instructor, which is the creation of in-class activities for the students. We give several examples of such activities, relating them to the research literature on teaching. Finally, in Section~\ref{sec:results}, we summarize our experiences teaching this course.

\section{Course Structure}
\label{sec:course-structure}
The course was introductory quantum field theory for particle physicists, usually taken by a mixture of first-year and second-year grad students and a very small number of advanced undergrads. Class size averaged around 20~people in all. It met twice a week for 70 minutes in a classroom with separate tables for groups of four or five students each, surrounded by blackboards. This course was taught four times in four consecutive years.

The instructor began the course with an explanation of the course format and why it was being used. The students were told that this was a “reading course,” and that as practicing physicists it would be important to know how to learn on their own by reading scientific literature. They were told that there was much in the course that they would learn just from the readings, material that would not be repeated in class but which they were responsible for. Class time was reserved for issues raised in the readings that were particularly subtle or important, or easily misunderstood. These would be addressed collaboratively.

The following were the main components of the course:
\begin{itemize}
    \item \textit{Small number of key themes:} The course had a coherent story line throughout, with a small number of ideas/questions appearing again and again in different parts of the course: for example, Where do quantum field theories come from? What information goes into the design of a field theory? How do we extract experimentally useful information from such a theory? What techniques are available, how do we use them, and what are their limitations? How do we use symmetries to extract useful information? 

    \item \textit{Assigned reading for every class meeting:} Students were assigned about 10~pages of reading per class meeting, which were mostly from a textbook\cite{peskin} but in some parts of the course ventured into the research literature. The lecturer posted the reading assignments online several days before class, introducing each assignment with a paragraph or two of text that highlighted what was important in the reading and how it related to the course as a whole. 

    \item \textit{Online questions about reading:} Students were asked to answer two to four questions on the reading and post their answers online the night before class, for example, by uploading cellphone photos of their handwritten answers. The questions took students roughly 10--30~minutes to finish, after doing the reading. It took the instructor about 30~minutes to go through the students' answers each time, and their answers would usually affect what was done in class the next day. The students were told that the questions were the sort of questions that professional physicists, such as the instructor, would ask themselves as they read these materials, to test their understanding.\cite{homework} Sometimes a question would focus on a detail in a particular paragraph or equation, when it was important that everyone look closely at that paragraph/equation. Other times a question might prepare students for a discussion planned for the next class. The last question was always a variation on “What about the readings needs more explanation?” This was frequently illuminating for the instructor, even after teaching the course multiple times.

    \item \textit{In-class discussion of reading:} The instructor start\-ed each class by commenting on students’ answers to the online questions, usually referencing specific answers (anonymously). This was an opportunity to straighten out misunderstandings and/or to set up activities for the rest of the class period. It also underscored for the students that the reading and online questions were essential components of the course, fully integrated with the in-class  activities. 

    \item \textit{In-class worksheets:} The instructor next had the class work two to four problems on worksheets that were handed out. (Sample problems are discussed in Section~\ref{sec:activities}.) \red{The class worked on one problem at a time.} Typically the instructor introduced the problem with a few minutes of lecturing, and then had students work on it. They first worked individually, and then discussed their results with their group of four or so students seated at a table. While this went on, the instructor circulated through the room, looking at the students' work \red{(on their worksheets)} and possibly discussing it with them. The instructor frequently failed to get through all of the prepared problems; the remaining problems often ended up on homework \red{or as online questions.}
    
    The instructor stopped each exercise once most students had completed most or all of the problem. He then described how an expert would approach the problem, usually referencing work the students had done individually and sometimes calling on individuals or groups to explain. In all, the instructor lectured for as much as~50\% of the time in class, depending on the topic. Some lecturing is \textit{essential} to active learning,\cite{schwartz-bransford, chapterJ} but most of the lecturing came in pieces after students had tried out the concepts themselves on the worksheets\,---\,they had to earn the lecture. Students handed in their (signed) worksheets at the end of class, to claim participation credit for their work; the worksheets were returned at the next meeting. The instructor would sometimes examine the worksheets before they were returned, but this was usually unnecessary because he had seen them while they were being filled out (and he was circulating).

    \item \textit{Worksheet solutions:} After every class the instructor posted his solutions to the worksheet problems online. These functioned as lecture notes for the course.

    \item \textit{Homework and solutions:} Students were assigned substantial homework assignments every one to two weeks, as in a typical graduate course. These were followed by detailed solutions, posted online by the instructor the day after the homework was handed in (while students still remembered the assignment). The solutions provided the instructor an opportunity to explain how the homework exercises related to the larger themes of the course.
\end{itemize}

The grading scheme for the course was designed to provide feedback to students on their performance and to signal which activities were important. \red{Students received feedback soon or immediately after they completed the online questions and in-class worksheets, since the instructor discussed these in class. They were made accountable for this work, by handing it in to the instructor, and they received participation credit (15\%) for it in their final grade; it was not graded for correctness.} The bulk of the final grade was awarded for correct answers on the homework~(60\%) and on a take-home, open-book final exam~(25\%).

\section{In-Class Activities}
\label{sec:activities}
A lecture in a traditional course is usually built around two or three key ideas\cite{overload} that the instructor wants students to take away from the class. In an active learning class, the instructor designs two or three activities around the same ideas.\cite{optics} The activities are designed to help students understand what issue is resolved by each idea and how it is resolved\,---\,for example, how time-ordered operators allow us to extract physics (i.e., $S$-matrix elements via the LSZ Theorem) from quantum field theories. Importantly, these activities also allow students to try out ideas with the instructor present, to give them immediate feedback on whether they are getting the point. 

There is much research showing that student learning is greatly enhanced, often by factors of two or three \red{(as measured in diagnostic tests)}, if students struggle with a problem \emph{before} they are told how to solve it.\cite{schwartz-bransford, chapterJ} There is also evidence that students who have learned concepts this way are better able to transfer those concepts to novel contexts.\cite{transfer1, transfer2} So when teaching a key idea, an instructor should not start by explaining the idea and then giving students examples to practice on. Instead, the instructor seeks to do the reverse: first give students an example that challenges them to discover the key idea themselves, and then give them a mini-lecture that shows them how an expert would organize the problem. This suggests a strategy for selecting problems: decide which topics warrant a mini-lecture, and then design a student activity that leads into each mini-lecture.

Research also shows that there is great advantage in having students work problems in small groups (3~or 4~people). Students learn at least as much from discussing a problem with other students as they do from hearing the instructor lecture on the problem,\cite{groups} even when everyone in the group is wrong.\cite{groups2} Group work benefits weaker and stronger students alike, unlike lecturing which appears to be particularly ineffective for stronger students.\cite{groups} But the combination of small group discussions followed by a mini-lecture is substantially more effective for everyone than either separately.\cite{groups} 

The unexpected power of student-student social learning is a major result from education research. This works partly because groups provide feedback for students when the instructor is elsewhere, but it is also because the brain works differently when teaching or critiquing a peer than when thinking about a topic alone. Critiquing is a particularly important expert skill that students typically have little experience with. Working in groups also helps keep individuals from getting stuck and makes it easier for the instructor to monitor student progress (e.g., by listening in on their conversations).

The following sections describe sample activities used in the quantum field theory course.

\subsection{First Problem}
The first activity in the course occurred in the first meeting, before students had done any reading. Their exercise concerned the origins of the wave equation that describes waves in an infinite stretched string. They were to derive the wave equation ``the way a theoretical physicist would derive it,'' using symmetries (but \emph{not} using Newton's laws). The instructor did the first part. He used locality and translation invariance (in $x$ and $t$), together with the assumption of small amplitudes, to argue that the general form of the equation describing displacements $y$ of the string from equilibrium is
\begin{align}
    0 &= b_0 + b_1 y + b_2\frac{\partial y}{\partial x} 
    + b_3\frac{\partial y}{\partial t}
    \nonumber \\
    &+b_4\frac{\partial^2 y}{\partial x^2} 
    + b_5 \frac{\partial^2y}{\partial x\partial t}
    + b_6\frac{\partial^2 y}{\partial t^2}
    \nonumber \\
    &+b_7\frac{\partial^3 y}{\partial x^3} + \cdots
\end{align}
with derivative terms of arbitrarily high order.
There are also terms nonlinear in~$y$ and its derivatives, but these were \red{neglected initially.}
The students were then asked to work in groups on the following questions:
\begin{quote}
    \textbf{Question 1:} Identify symmetries of the stretched string that remove further terms from the general equation (above). There are  at least four, although you may not need all of them. Characterize each symmetry using the ``If $y(x,t)$ is a solution, so is\ldots'' formulation. Also write a sentence describing the symmetry.
\end{quote}
\begin{quote}
    \textbf{Question 2:} The wave equation is buried in the list of 
    possible terms. In what limit(s) would the two wave-equation terms dominate the others in this list? Are there restrictions on the relative sign of the the two wave-equation terms?
\end{quote}
\begin{quote}
    \textbf{Question 3:} What are the most important corrections to the wave 
    equation given your approximation(s) (and consistent with all of 
    your symmetries).
\end{quote}
The second and third questions were there to challenge  groups
who got through the first question more quickly than the rest of the class. 

Most groups came up with time-reversal symmetry (if $y(x,t)$ is a solution, so is $y(x,-t)$), which rules out terms with an odd number of derivatives~$\partial/\partial t$, and space-reflection symmetry or parity, which rules out terms with an odd number of derivatives~$\partial/\partial x$. Usually only one or two groups would think of $y$-displacement symmetry (if $y(x,t)$ is a solution so is 
$y(x,t) + \Delta$ where $\Delta$ is a constant). This symmetry, which rules out 
the $b_1y$~term, is a classical example of spontaneous symmetry breaking and Goldstone's Theorem, which played a significant role later in the course.
The constant~$b_0=0$ because $y(x,t)=0$ is the equilibrium solution, but can also be ruled out by a $y\to-y$ symmetry which many groups proposed. Occasionally groups would suggest more complex symmetries, like variations on a dilatation symmetry.

Symmetry and stability arguments reduce the general equation to something that can be written\cite{time-derivatives}
\begin{align}
    \frac{1}{c^2}\frac{\partial^2y}{\partial t^2}
    = \frac{\partial^2y}{\partial x^2} 
    + g_2 a^2 \frac{\partial^4y}{\partial x^4} 
    + g_4 a^4 \frac{\partial^6y}{\partial x^6} + \cdots
\end{align}
where $c$ is the wave speed, $a$~is a length that is somehow related to physical characteristics of the string (e.g., the string's diameter), and the $g_{2n}$~are dimensionless. The second question asks under what circumstances the $a^{2n}$~terms in this equation can be neglected. Most groups found this difficult to answer. The instructor helped individual groups move forward by suggesting that they write down solutions of the wave equation and substitute them into the equation. Eventually they figured out that the $a^{2n}$~terms are suppressed, by $(a/\lambda)^{2n}$, for solutions~$y(x,t)$ that involve only long wavelengths~$\lambda \gg a$\,---\,leaving behind the 
wave equation.

The wave equation is not exact in this analysis, but has a hierarchy of correction terms. Terms with many powers of~$a^2$ are more suppressed than terms with few or no powers. The number of powers is determined by the dimension of $\partial^ny/\partial x^n$. The third question focuses attention on the leading correction, whose form is completely specified by symmetries up to a single unknown constant~$g_2a^2$. 

\red{This analysis shows how the wave equation (with its corrections) emerges from very general assumptions about locality and symmetries, as a low-amplitude, long- wavelength, low-frequency approximation to other arbitrary dynamics. These properties are not uncommon, and this is why the wave equation arises in so many different contexts. It represents the \textit{universal} behavior of systems with these (quite generic) characteristics.

These ideas carry over to quantum field theory as was emphasized by closely related activities in the subsequent days and weeks:}
\begin{itemize}
    \item Students were asked to derive the Lagrangian, including higher-dimension correction terms, for a scalar field theory using Lorentz invariance, parity symmetry, etc., together with a low-energy/momentum expansion. This yields the Lagrangian used to introduce quantum field theory in the early weeks of the course. 
    
    \item Dimensional analysis of the scalar theory indicates that the scalar particle's mass is naturally very large\,---\,so large that a low-energy expansion is not possible, thereby invalidating the Lagrangian's derivation.\cite{nonrel} Students were asked what symmetry could be used to address this issue and what the resulting Lagrangian is. 
    That symmetry is $\phi\to\phi+\Delta$, in analogy to the wave equation symmetry that gets rid of the $b_1 y$~term. Mass terms violate this symmetry, so
    the resulting theory describes massless scalar particles.\cite{weakinter} This theory is a useful prototype for the chiral theory that describes low-energy pions, and also the theory that describes the (hypothetical) axion particle. \red{ It is also the first time students confront the ``mass problem'' in quantum field theory, which is a primary driver for current experimental work in particle physics.}

    \item Students were asked to use gauge symmetries, Lorentz invariance, etc.\ to derive the QED Lagrangian, including corrections, first for charged scalar particles, then for neutral scalars, and later for charged and neutral spin-$1/2$ particles. These Lagrangians were central to all discussions in the last half of the course. The theories describing the QED interactions of neutral particles are not in standard textbooks, \red{which makes them particularly useful for activities.}
\end{itemize}

The ideas and techniques applied in the wave equation problem, on day one, are fundamental to the modern understanding of the Standard Model of particle physics as a low-energy/momentum effective field theory.\cite{renormalization} Introducing these ideas early on gave students an opportunity to play with them before they had to grapple with the complexities of quantized fields, Dirac equations, renormalization, and all the other topics in the course. Revisiting them repeatedly throughout the course gave students an opportunity to practice the ideas in different contexts, thereby exercising and updating the mental maps they use to organize material from the course.\cite{expert-novice} Many students had heard of things like spontaneous symmetry breaking and axions. Having these appear early in the course helped reinforce the course's relevance to their larger interests.

\removefig{
\begin{figure}
    \begin{center}
        \includegraphics[scale=0.95]{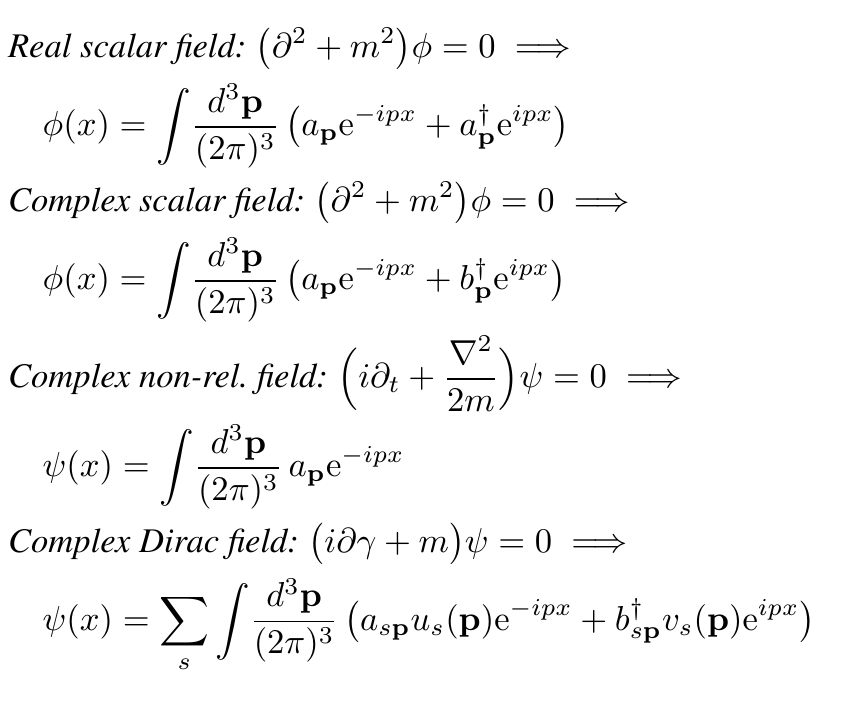}
    \end{center}
    \caption{\label{fig:contrasting-cases} Contrasting cases illustrating when a quantum theory requires a new anti-particle, distinct from the particle.}
\end{figure}
}

\subsection{Contrasting Cases\cite{daed2}}
\red{The following example illustrates the use of  \textit{contrasting cases}  for learning general principles.\cite{clowns, chapterC} Anti-particles appeared early in the course. Rather than explaining the rules governing the introduction of anti-particles, the instructor gave the students several examples or cases and challenged them to find the rules:}
\begin{quote}
\textbf{Question:} One of the early triumphs of relativistic quantum mechanics was its prediction of the existence  of anti-particles. Consider the four types of  (non-interacting) quantum field theory
shown in Fig.~\ref{fig:contrasting-cases}.
The second and fourth theories require new anti-particles, distinct from the original particles. The others do not. What is the general rule that determines
whether there is a new anti-particle?
\end{quote}
The students had read about 
the first two cases in the textbook, and were familiar with the third case 
from previous worksheet exercises. 
Given only the first two cases, one might conclude that the general rule is that complex-valued fields have new anti-particles. That idea is undermined, however, by the third example, which has a complex field but no anti-particle. Maybe the third case is special because its field equation is linear in~$\partial_t$, unlike the first two which are quadratic. The fourth case, however, also has a field equation that is linear in~$\partial_t$, but requires a new anti-particle.\cite{majorana}

\red{The actual rule has two parts: one needs to determine both whether there are negative energy solutions to the field equations,\cite{neg-energy} and, if so, whether the number of degrees of freedom encoded in the field requires independent Fourier coefficients for the positive and negative-energy terms.\cite{dof}. In practice, many students managed to understand the relevance of the number of degrees of freedom, but few paid attention to the need (or not) for negative-energy states, despite this being the starting point for most elementary discussions of anti-particles\,---\,this was prior knowledge that needed to be activated.} 

\red{As discussed above, having students struggle to find a rule is far more effective than telling them the rule and having them practice on examples.\cite{clowns, chapterC}} The contrasting cases were carefully selected to make it difficult for students to fasten on a simpler rule by thinking too narrowly about the possibilities. Each of the surface features mentioned above\,---\,complex versus real, first-order versus second-order, scalar versus matrix\,---\,is relevant to the correct rule, but none is definitive on its own. A major point of the exercise is to get students to think through how these surface features relate to the real issues. This illustrates how the same principles of learning (e.g., distinguishing key underlying structures from surface features when approaching a new problem) apply across the physics curriculum from introductory mechanics to quantum field theory.

\removefig{
\begin{figure}
    \includegraphics[scale=0.95]{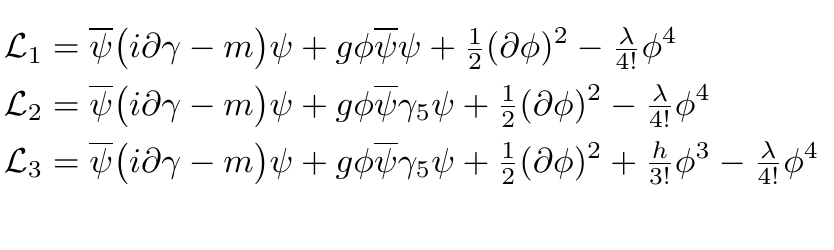}
\caption{\label{fig:multiple-choice} Lagrangians that illustrate the 
systematics of parity invariance.}
\end{figure}
}

\subsection{Multiple-Choice Review\cite{daed2}}
\red{Research strongly suggests that review activities have a much larger impact on student learning than review lectures, even if students have trouble recalling the earlier lessons.\cite{chapter1,rev-quiz} This is especially true of in-class activities since they force students to rely upon their memories rather than texts or notes.} Getting information into the brain is relatively easy; it is getting it out again that is difficult and requires effortful practice. The  interconnections in the brain that allow for retrieval are strengthened by such practice, just as muscles respond to exercise\,---\,the brain is physically modified.\cite{brain-modified}

\red{This problem was given as an in-class activity to review how parity is expressed in quantum field theories:}
\begin{quote}
    \textbf{Question:} Consider a spin-1/2 spinor field~$\psi$ coupled to a real-valued spin-0 field~$\phi$ in each of the three Lagrangians 
    in Fig.~\ref{fig:multiple-choice}.
    For each case indicate whether the Lagrangian has a symmetry under a parity transformation, remembering that $\psi(\xv,t)\to\gamma^0 \psi(-\xv,t)$ under parity. Choose one of:
    \begin{itemize}
        \item [a)] Yes % \hspace{0.25\columnwidth} b) No
        \item [b)] No 
        \item [c)] It depends on whether $\phi$ is a pseudo-scalar\\ or a scalar.
    \end{itemize}
\end{quote}
The techniques used to analyze spin-1/2 parity are mechanical, and so were relegated to out-of-class reading and homework. The review \red{activity,} coming after the reading, provided an opportunity to discuss the techniques in class to whatever extent made sense \red{given how the students, individually and in their groups, approached the problem.}

An additional reason for this problem was a widespread and surprisingly durable misconception that was apparent every year in teaching this course. Students tend to think of a field’s behavior under parity (i.e., whether it is a pseudo-scalar or a scalar) as a property intrinsic to the field. In fact, it is a property of the Lagrangian of the theory: the field’s transformation under parity is whatever it takes (if anything) to make its Lagrangian invariant when the spatial coordinates are reflected through the origin ($\xv\to-\xv$). 

\red{The correct answers for this problem are~a), a), and~b).\cite{parity} 
Despite earlier readings and practice problems, and explicit discussions in earlier classes,  a large fraction of the
class chose~c), c), and~c), and almost no one had a perfect score. Answer~c) is nonsense because it ignores the underlying Lagrangian. Multiple-choice problems can help students appreciate that they harbor misconceptions because the problems can be constructed\,---\,here, by including option~c)\,---\,to guarantee that everyone engages with the misconception.}

Multiple-choice problems also leave no question in the students’ minds about whether they understand the underlying idea\,---\,if they chose~c) they didn't understand. It catches their attention. One student leaving class was overheard to comment wryly: “Three out of three wrong. This is the worst physics quiz result I have ever received.” 

\subsection{Other Strategies}
\label{sec:other-strategies}
A variety of approaches were used to generate problems for in-class group work, always with the requirement that they would engage students in the skills and reasoning processes a theoretical physicist would routinely use in their work:
\begin{itemize}
    \item Immediately after students read about a standard technique, they would often be asked to apply the same technique to a different field theory, one not discussed in the text. For example, such methods as canonical quantization of fields, and the Schwinger-Dyson equation for generating perturbation theory were introduced in the text using a relativistic scalar field theory.  The students were asked to apply these methods to a non-relativistic theory, which is sufficiently different to be nontrivial but simple enough to be doable in class.

    \item In-class exercises were useful for giving students practice with the mechanics of new techniques. Their progress can be slowed by trivial mistakes and misunderstandings that are readily resolved in the classroom. For example, when first assembling a Feynman-diagram contribution that involves spin-$1/2$ particles, students frequently wrote things like
    \begin{align}
        \label{eqn:wrong}
        (\slashed{k} + m)\;ie\gamma_\mu\; \overline{u}(p)u(q)\;ie\gamma^\mu
    \end{align}
    rather than
    \begin{equation}
        \overline{u}(p) ie\gamma_\mu(\slashed{k} + m)ie\gamma^\mu u(q).
    \end{equation}
    They were forgetting that $\gamma_\mu$, $\slashed{k}$, and  $m$ all represent $4\times4$~matrices, while $u(q)$ and $\overline{u}(p)$ are column and row 4-vectors, respectively\,---\,so the order in which you write them down matters. That Eq.~(\ref{eqn:wrong}) is nonsense is obvious to an expert, but less so to a novice. Such simple misunderstandings are addressed very efficiently in the classroom, where an expert is close at hand.

    \item \red{In-class activities are particularly well suited to connecting course content with  experimental results because groups are more likely to succeed in making the connections than individuals. Also it creates an opportunity for the instructor to elaborate on the larger signficance of the results immediately after the students have engaged with them.} Late in the course, for example, the students worked through the consequences of spontaneous symmetry breaking for a simple Yukawa theory where a spin-$1/2$ particles is coupled to a neutral scalar. They were then asked how their analysis was related to a famous plot from CERN which shows the quark-Higgs coupling as a function of quark mass. The model the students analyzed is a toy model, not the real Standard Model Lagrangian, and so was sufficiently simple to be studied in class. It nevertheless exhibits the correct relationship between the coupling and quark mass, and so provides a qualitative explanation of the experimental data. \red{This experimental result was important evidence, when it came out, that quark masses (and probably all other masses) come from spontaneous symmetry breaking\,---\,a major advance.}

    \item To test their understanding of a derivation in a reading assignment, students were sometimes asked to explain precisely what goes wrong with the derivation if one of the assumptions is changed. For example, students were asked how the derivation of the conserved energy-momentum tensor for a scalar theory was invalidated if the particle's mass depended on time. A variation on this assignment would be to ask students to invent a Lagrangian for which the energy-momentum tensor is not conserved.

    \item Students were asked to connect their current work to ideas from previous weeks. \red{Again this works particularly well as an in-class activity because of the groups and timely instructor feedback.} When deriving representations of the Poincar\'e group for spin-$1/2$ particles, for example, students were asked what equation for massive spin-$1$ particles corresponds to the Dirac equation. The corresponding equation for spin-1 is $\partial^\mu A_\mu = 0$. These equations are needed because the fields in both cases have more components than there are spin states. This connection is not obvious to most students, but reflects a fundamental tension in quantum field theory between locality and Lorentz invariance. \red{An expert's conceptual map of the subject is built out of these kinds of insights.}

\end{itemize}

\section{Results and Conclusions}
\label{sec:results}

Students were generally open to active learning, though it took time for some to get used to the group work. Attendance in class was nearly perfect (95\%), despite the 8:30\,am start time.\cite{attend} End-of-semester student evaluations of the course were mediocre~(3.7/5) in the first year, but improved substantially as the course evolved, until they were as positive as they could be~(5/5) in the last year.\cite{vg} The instructor circulated throughout the classroom continuously while students worked on activities, helping groups both with the physics and with group dynamics. This informal contact meant that discussions following an activity  were far more lively than in a traditional lecture, with a much larger fraction of the class participating. There was also substantially more participation in after-class discussions and office hours than in earlier courses taught traditionally by the instructor.

One byproduct of the larger role for students in the classroom was that teaching the course never became stale or boring for the instructor, no matter how often he taught it\,---\,each new group of students brought fresh perspectives and raised new issues that often changed the course of class meetings. It also gave the instructor more insight into the obstacles encountered by students. Some of these involved incorrect preconceptions about quantum field theory, as discussed earlier. Others involved tools they had been taught in previous courses. While they understood the physics behind these tools, they had insufficient practice with them to attain fluency.\cite{fluency} This lack of fluency was evident in the in-class activities and led to changes in their designs in subsequent years, as well as to new online questions and homework problems that provided extra practice.

Course coverage was not compromised by active learning. This was because straightforward topics were relegated to the reading assignments and not covered in class. Students did the reading, as was evident from their responses to the reading questions submitted online before each class meeting\,---\,over the semester, students earned on average~96\% of the available participation credit for the online questions.\cite{heiner} The material covered by the course was fairly standard for an introductory quantum field theory course, covering roughly the first 200--250~pages of standard texts,\cite{peskin} together with a discussion of renormalization theory through leading-log order built on material from Ref.~\onlinecite{renormalization}. Homework assignments and the final exam were very similar to what is used in conventional courses, and the students did well on both (average and standard deviation on the final were~72\% and~13\%, respectively).

Moving topics to the reading assignments also freed up time to talk about the course and the current state of quantum field theory and particle physics. Quantum field theory textbooks, for example, are highly optimized in their choice of topics, notation, and emphasis for applications of weak-coupling perturbation theory (Feynman diagrams). This reflects the predominance of perturbation theory in the history of the subject. Nonperturbative techniques, like the numerical methods used in QCD simulations or the mathematical methods used in conformal field theory, have become far more important in recent decades. It was a useful (metacognitive) exercise to imagine how textbooks and courses might be different if they were optimized for nonperturbative analyses.\cite{composite}

\red{The history of quantum field theory is itself illuminating about the importance of practice with expert feedback when learning a hard subject.} David Kaiser, in his book \textit{Drawing Theories Apart}, traces the dispersion of Feynman-diagram techniques through the theoretical physics community in the first several years after Feynman invented them. Kaiser finds that the only theorists who adopted Feynman diagrams in that period were those who had had extended \emph{personal} contact with Feynman or Feynman's {former Cornell colleague} Freeman Dyson, or with one of their postdocs or graduate students. This is despite the fact that Feynman's techniques were dramatically more efficient than the alternatives, and were well documented in the literature. Kaiser summarizes:\cite{Kaiser}
    \begin{quote}
        ``No uncrossable, epistemic barriers separated diagram users from nonusers. Physicists could certainly learn \textit{something} about the diagrams from the published literature or from Dyson's unpublished lecture notes. Yet reading texts is not the same as using tools. Long after textual instructions became widely available, almost no one used the diagrams in actual calculations upon learning about them from texts alone\ldots Feynman diagrams are \textit{practices,} and as such they must be \textit{practiced}\ldots''
    \end{quote}
And that practice entailed ``personal contact and sustained, face-to-face training'' from people who had already mastered the techniques.  

The main goal of a graduate quantum field theory course is to make novice quantum field theorists more expert. The expert knows more facts about quantum field theory, but, more importantly, also knows how to organize and connect those facts much more effectively. The expert knows how to choose the right tool for a new problem, how to evaluate and test solutions, and how to select/modify a problem so it is relevant and solvable. Short of one-on-one tutoring, active learning offers an instructor the most powerful tools available for helping novices become expert at such skills, since it allows students to practice the skills (in the classroom) with an expert close by. As discussed above, textbooks are poor substitutes for this kind of training; so are conventional lectures (in person or online) for much the same reason\,---\,little or no real-time feedback to students. Student engagement and enthusiasm are much higher in an active learning class. And the constant, close interaction with students makes active-learning classes  endlessly stimulating and enjoyable for the instructor. 

\begin{acknowledgments}
We thank Carl Wieman for his numerous comments on the manuscript and for years of advice on all matters connected with active learning.  
\end{acknowledgments}

\removefig{
\end{document} 
}

\newpage   % Start a new page for figure captions

\section*{Figure captions}

\begin{figure}[h!]
    %% original source:
    % \begin{center}
    % \begin{align}
    %     &\mbox{\textit{Real scalar field:}} 
    %     \; \big(\partial^2 + m^2\big)\phi = 0 \implies
    %     \nonumber \\
    %     &\quad\phi(x) = \int\!\frac{d^3\pv}{(2\pi)^3}\,\big(a_\pv \er^{-ipx} 
    %     + a^\dagger_\pv \er^{ipx}\big)
    %     \nonumber \\
    % % \end{align}
    % % \begin{align}
    %     &\mbox{\textit{Complex scalar field:}}
    %     \; \big(\partial^2 + m^2\big)\phi = 0 \implies
    %     \nonumber \\
    %     &\quad\phi(x) = \int\!\frac{d^3\pv}{(2\pi)^3}\,\big(a_\pv \er^{-ipx} 
    %     + b^\dagger_\pv \er^{ipx}\big)
    %     \nonumber \\
    % % \end{align}
    % % \begin{align}
    %     &\mbox{\textit{Complex non-rel.\ field:}}
    %     \; \Big(i\partial_t  + \frac{\nabla^2}{2m}\Big)\psi = 0 \implies
    %     \nonumber \\
    %     &\quad\psi(x) = \int\!\frac{d^3\pv}{(2\pi)^3}\,a_\pv \er^{-ipx} 
    %     \nonumber \\
    % % \end{align}
    % % \begin{align}
    %     &\mbox{\textit{Complex Dirac field:}}
    %     \; \big(i\partial\gamma + m\big)\psi = 0 \implies
    %     \nonumber \\
    %     &\quad\psi(x) = \sum_s\int\!\frac{d^3\pv}{(2\pi)^3}\,\big(a_{s\pv} u_s(\pv) \er^{-ipx} 
    %     + b^\dagger_{s\pv} v_s(\pv) \er^{ipx}\big)
    %     \nonumber
    % \end{align}
    % \end{center}
    %%
    % \includegraphics[scale=0.95]{lepageFig01}
    \caption{\label{fig:contrasting-cases} Contrasting cases illustrating when a quantum theory requires 
    a new anti-particle, distinct from the particle.}
\end{figure}

\begin{figure}[h!]
%% original source:
% \begin{align}
%     \mathcal{L}_1 &= \psib\big(i\partial\gamma - m\big)\psi
%     + g\phi\psib\psi + \tfrac{1}{2}(\partial\phi)^2   
%     - \tfrac{\lambda}{4!}\phi^4 
%      \nonumber \\
%     \mathcal{L}_2 &= \psib\big(i\partial\gamma - m\big)\psi
%     + g\phi\psib\gamma_5\psi + \tfrac{1}{2}(\partial\phi)^2   
%     - \tfrac{\lambda}{4!}\phi^4 
%     \nonumber \\
%     \mathcal{L}_3 &= \psib\big(i\partial\gamma - m\big)\psi
%     + g\phi\psib\gamma_5\psi + \tfrac{1}{2}(\partial\phi)^2   
%     + \tfrac{h}{3!}\phi^3- \tfrac{\lambda}{4!}\phi^4 
%     \nonumber    
% \end{align}
%% 
% \includegraphics[scale=0.95]{lepageFig02}
\caption{\label{fig:multiple-choice} Lagrangians that illustrate the 
systematics of parity invariance.}
\end{figure}


\begin{thebibliography}{99}
\bibitem{homework} \red{Several of the approaches described in Sec.~\ref{sec:other-strategies} are also useful for formulating online questions.}
\bibitem{time-derivatives} There is no need for terms involving extra time derivatives, like $\partial^4y/\partial t^4$, since these can be replaced by terms with only spatial derivatives by using the wave equation (for solutions). This is closely related to the topic of \textit{redundant operators} in a quantum field theory.
\bibitem{nonrel} {Scalar field theory makes sense as an effective field theory with a finite ultraviolet cutoff~$\Lambda$ of order the scale at which new physics, beyond the scalar theory, appears. Contributions from high-dimension operators (e.g., $g\phi^6/\Lambda^2$) are normally suppressed in applications by powers of $p/\Lambda$ where $p$ is a momentum typical of the application. The natural size for the mass term, however, is 
   $\Lambda^2\phi^2$, which means that: the $\phi$ particle's mass is of order~$\Lambda$, $p/\Lambda\approx1$, and high-dimension operators are not suppressed. A low-3-momentum expansion is still possible, leading to a non-relativistic effective field theory.}
\bibitem{weakinter} \red{The particles are also weakly 
   interacting because the leading interaction has dimension eight and so is suppressed by $1/\Lambda^4$: $\mathcal{L}_\mathrm{int} = - g (\partial\phi\cdot\partial\phi)^2/\Lambda^4$.}
\bibitem{majorana} Another example that could have been included is the Majorana spinor field, which is a relativistic theory with a complex field and a first-order field equation, but no new anti-particle. Students were given an opportunity to revisit the thinking used in the present problem when they were analyzing Majorana fields for a homework problem several weeks later.
\bibitem{neg-energy} Relativistic theories typically have both positive-energy and negative-energy solutions, because $E^2=\pv^2+m^2$ has two solutions. Such theories need two terms in the Fourier expansion: $\exp(-ipx)$ for the positive-energy solutions, and $\exp(ipx)$ for the negative-energy solutions, which are associated with the anti-particles. Non-relativistic theories have only positive-energy solutions (because $E=\pv^2/2m$ has only one solution), and so only one term. The last case, like the third case, is linear in $i\partial_t\leftrightarrow E$, but it is a matrix equation whose solution leads immediately to $E^2=\pv^2+m^2$.
\bibitem{parity} In the first case, the theory has a parity symmetry where the $\phi$~field transforms like a scalar under parity ($\phi(\xv,t)\to\phi(-\xv,t)$). Thus, $\phi$ is necessarily a scalar if it is described by this Lagrangian. Similarly, it is necessarily a pseudo-scalar in the second case ($\phi(\xv,t)\to-\phi(-\xv,t)$). The third Lagrangian breaks parity symmetry because $\phi^3$ changes sign when $\phi(\xv,t)\to-\phi(-\xv,t)$. So $\phi$ is neither a scalar nor a pseudo-scalar in this theory; parity is a useless construct here, and the pseudo-scalar/scalar distinction is meaningless.
\bibitem{attend} This is in part because of the participation credit given for in-class work. Credit towards final grades should signal what the instructor believes are essential components in the course. Students respond to these signals.
\bibitem{vg} This does not mean, obviously, that the course is perfect. There remain many opportunities for significant improvement.
\bibitem{composite} A nonperturbative focus, for example, would spend time on composite particles (bound states). QCD's physical states are all composite, but most field theory texts present no treatment of such states.
\end{thebibliography}
\end{document}